\documentclass[a4paper,11pt]{article}
\usepackage{jinstpub} 
\usepackage{lineno}
\usepackage{enumitem}
\usepackage{booktabs} 
\usepackage{xspace} 
\usepackage{wasysym}

\title{GPU-Accelerated Simulation of 3D Diamond Sensors Using the TeRABIT Infrastructure}

\newcommand\comsol{COMSOL MultiPhysics\textsuperscript{\tiny\textregistered}\xspace}

\author[a]{L.~Anderlini}
\author[*,a,b]{C.~Buti}
\author[a,b]{E.~Eredi}
\author[a]{G.~Passaleva}

\affiliation[a]{INFN Firenze,\\ via G. Sansone 1, Sesto Fiorentino, Italy}
\affiliation[b]{University of Florence, Physics and Astronomy Department,\\ via G. Sansone 1, Sesto Fiorentino, Italy}
\emailAdd{Clarissa.Buti@fi.infn.it}

\abstract{
Diamond detectors with electrodes orthogonal to the surface, engraved via laser-induced graphitization, are full-carbon sensors of interest for a wide range of applications, spanning from High Energy Physics to Nuclear Medicine and dosimetry.
In recent years, significant progress has been made in graphitization techniques, enabling the fabrication of lower-resistance electrodes. This has resulted in faster sensors, achieving time resolutions better than 100 ps.
However, simulating signal formation in these devices remains a challenge. The effects of fluctuations in energy deposition, carrier transport, signal propagation, and readout electronics intertwine in a way that is non-trivial to disentangle.
We have developed an innovative simulation approach based on an extension of the Ramo–Shockley theorem, modeling propagation effects in a theoretically sound manner by introducing time-dependent weighting potentials. These are obtained by solving a third-order partial differential equation derived as a quasi-static approximation of Maxwell’s laws. The numerical solution of this equation emerged as the main challenge of the new approach.
In this contribution, we discuss an innovative solver that uses fundamental solutions to impose boundary conditions and spectral methods to extend the solution to the bulk of the diamond detector. We report on how the solver has recently been ported to GPUs and distributed across multiple computing sites, leveraging the TeRABIT HPC Bubbles and the InterLink protocol. This drastically reduces time-to-insight and effectively enables what-if studies on sensor geometry.
}


\begin{document}
\maketitle
\flushbottom
\section{Introduction}
Diamond is a large band-gap semiconductor studied for radiation detection due to its radiation hardness, low leakage current, solar-blindness, and ability to operate without cooling~\cite{Pomorski2009,pernegger}. Femtosecond laser pulses induce a phase transition from diamond to a mixture of graphite and amorphous carbon, enabling the fabrication of 3D full-carbon detectors with graphitized electrodes, which are relevant for dosimetry, beam-monitoring applications, and diamond-based active targets~\cite{Oliva:2019alx, KANXHERI202273,Anderlini:2021pei,Bachmair:2015iba}.

Due to the presence of amorphous carbon, the resistivity of graphitized structures is significantly higher than that of pure graphite, leading to large electrode resistance and limiting the achievable time resolution. In 3D diamond detectors with graphitized electrodes, the time resolution is affected by several factors, including fluctuations in charge deposition, carrier transport, and electronic noise; the dominant contribution arises from signal propagation along resistive electrodes.

This impedance effect is not accounted for in the standard Ramo-Shockley theorem; however, the extension developed by Riegler in the 2000s incorporates it, describing the induced current \( i(t) \) as a convolution between the velocity \( \vec{v}(t) \) of the carriers and a four-dimensional correlation function \( \vec{H}(\vec{x}, t) \). This function represents signal delay caused by impedance. In formula,
\begin{equation}
    \label{eq:ramoshockleyriegler}
    i(t) = -\frac{q}{V_w}\int_0^t \vec{H}(\vec{x}(t'), t-t') \cdot \vec{v}(t') \, dt',
\end{equation}
where q is the charge of the carrier in the sensor and \( V_w \), a normalization constant in volts, is conventionally set to 1 V~\cite{Gatti1982, 
    Riegler2019,
    Janssens:2890572}.

In our previous work~\cite{Anderlini:2025ffy}, we validated a numerical method using discrete Fourier transforms for computing \( \vec{H}(\vec{x}, t) \), comparing it with \comsol. The proposed pseudo-spectral method was found to be less computationally intensive than \comsol, but the overall simulation was still too expensive to perform systematic optimizations of geometrical parameters. The performance limitations stemmed from the architectural design of the software application.

The correlation function \( \vec{H}(\vec{x}, t) \) was computed in Python (\emph{numpy}), with performance-critical tasks implemented in custom C functions. While multi-processing and multi-threading were considered, they provided limited benefits, achieving at most a \( 4 \times \) speed-up.

The simulation of carrier trajectories within the diamond bulk was handled by Garfield++, which also performed the convolution described in Eq.~\ref{eq:ramoshockleyriegler}. However, this step was computationally expensive and could not effectively exploit parallelism, due to both memory constraints and the architectural design of Garfield++. 

In addition, the requirement of large memory allocations for representing  \( \vec{H}(\vec{x}, t) \) made it unsuitable for distributed computing on High-Throughput Computing (HTC) sites within the Worldwide LHC Computing Grid (WLCG), as these systems enforce strict per-core memory limits.

This work reports the porting of performance-critical simulation steps to GPUs and the distribution of payloads via the recent infrastructure by the TeRABIT project. The milestones include:
\begin{itemize}[noitemsep, topsep=0pt]
\item Migration to an orchestrated Directed-Acyclic-Graph (DAG) workflow (Section~\ref{sec:snakemake}).
\item GPU-acceleration of correlation functions and convolutions (Section~\ref{sec:weightingtide}).
\item Distributed computing combining INFN-CNAF (HTC) and TeRABIT Padova (HPC) resources (Section~\ref{sec:interlink}).
\end{itemize}
These advancements enable systematic scans of geometrical parameters, leading to a new sensor geometry optimized for timing (Section~\ref{sec:results}).

\section{Orchestrated workflow of micro-steps}
\label{sec:snakemake}

\begin{figure}
    \centering
    \begin{minipage}{0.2\textwidth}
        \caption{\label{fig:snakemake}\null
            Simulation workflow splitting CPU and GPU workloads. Logical and physical data flows are also shown. The logos of the underlying software frameworks or applications refer to, in order from the top to the bottom: Deuxfleurs Garage, CuPy, Garfield++ and Numpy.
        }
    \end{minipage}
    \hfill
    \begin{minipage}{0.7\textwidth}
        \includegraphics[width=\textwidth]{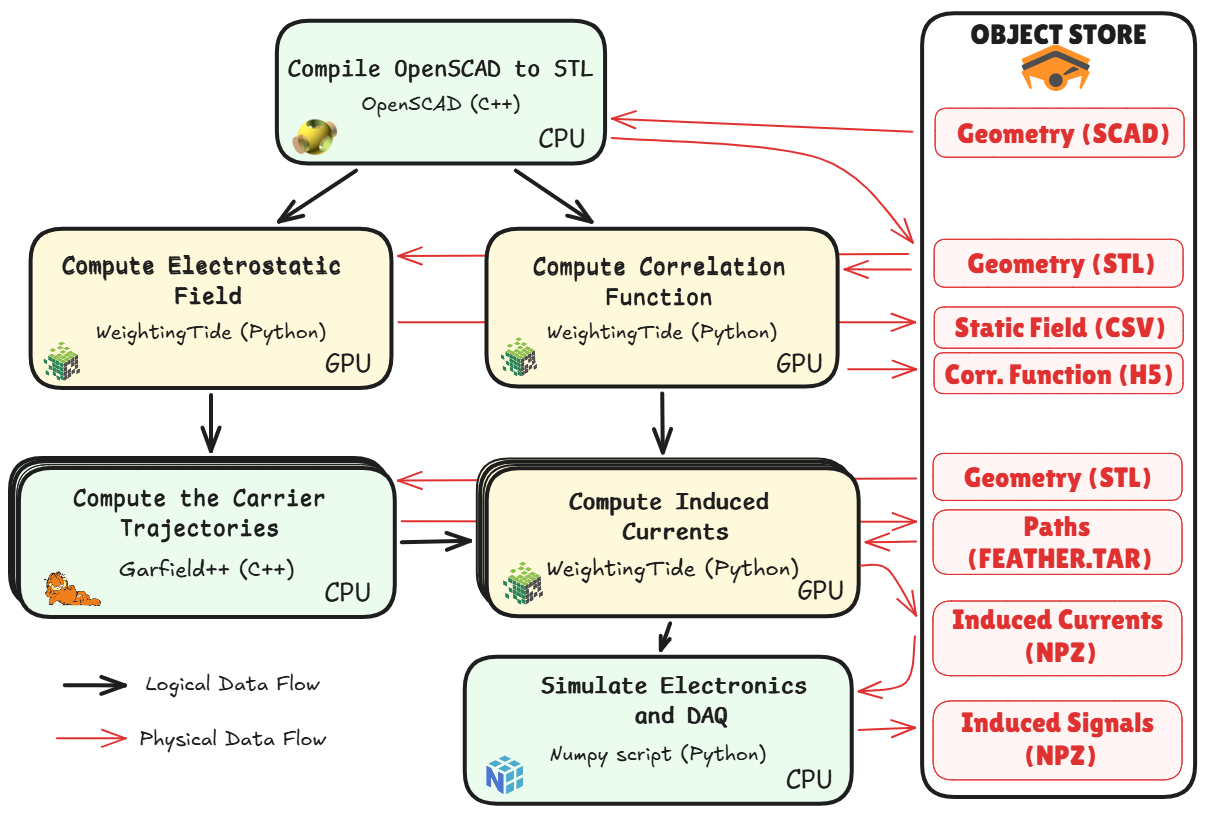}
    \end{minipage}
\end{figure}

The workflow is organized as a sequence of small steps, shown in Figure~\ref{fig:snakemake} and orchestrated by Snakemake~\footnote{see \href{https://snakemake.readthedocs.io/en/stable/}{snakemake.readthedocs.io}}. The workflow takes as input a text file describing the geometry of the conductive elements in the diamond and generates a set of 1000 signals corresponding to impinging 180~GeV/$c$ pions traversing the diamond specimen along the $z$-axis. 

The six steps to connect these two ends are described below.
\begin{itemize}
    \item \textbf{Compile OpenSCAD to STL}. The geometry is defined in the open format \emph{OpenSCAD}~\footnote{see \href{https://openscad.org}{openscad.org}} and is converted into Standard Tessellation Language (STL) by running the OpenSCAD via its command-line interface.

    \item \textbf{Compute Electrostatic Field}. The STL geometry is voxelized, to represent the sensor geometry in a regular three-dimensional grid. An iterative pseudo-spectral method, implemented using the CUDA wrapper of Python in CuPy~\footnote{see \href{https://cupy.org}{cupy.org}} to enable GPU hardware acceleration, is used to solve the Laplace equation in the diamond sensor. The output is stored in the comma-separated-value (CSV) format expected as input by Garfield++.
    
    \item \textbf{Compute the Correlation Function $\vec H(\vec x, t)$}. The computation of the correlation function is conceptually symmetrical to the computation of electrostatic field. The STL geometry is voxelized and a pseudo-spectral method is used to solve a differential equation. However, since the resulting maps are time-dependent and compatibility with legacy code is not a requirement, the binary H5 data format was chosen.

    \item \textbf{Compute the carrier trajectories}. The radiation-matter interaction of the simulated 180~GeV/$c$ pions with the diamond material is simulated using the HEED as implemented in Garfield++. Garfield++ is used to compute charge-carrier trajectories in the precomputed electrostatic field. The output consists of large CSV files (tens of MB per incident pion) describing individual carrier paths. To reduce network and storage overhead, the data are immediately converted to Apache Arrow format before the next processing step. This reduces file size by about 80\% and, more importantly, minimizes deserialization overhead in the subsequent GPU-accelerated correlation analysis.

    \item \textbf{Compute the induced currents}. For each charge carrier, the induced signal is computed as described in Equation \ref{eq:ramoshockleyriegler}. The convolution is offloaded to GPU using CuPy. The signals obtained from the convolution represent the current induced on the electrodes by the traversing particle and they are organized in a Numpy archive of a few kB.

    \item \textbf{Simulate Electronics and DAQ}. The final step (before the statistical analysis which is not included in the workflow) consists of convolving the induced signals with the transfer function of the readout electronics and adding the expected noise. These steps were implemented in a Jupyter notebook executed as the last step of the workflow.
\end{itemize}

The heterogeneity of the intermediate data formats is partially due to performance considerations, or to enable interfacing with existing software (such as OpenSCAD and Garfield++), but in some cases it is due to independent choices by the contributing physicists, usually driven by previous experience.

The steps based on CuPy were developed on purpose for this use-case and were collected in a Python package named \emph{WeightingTide} and discussed briefly in Section~\ref{sec:weightingtide}. The source code is publicly available in a GitHub repository~\footnote{see \href{https://github.com/landerlini/weightingtide}{github.com/landerlini/weightingtide}}.

To enable distribution of the payload, an S3-compatible storage provided by a Deuxfleurs Garage instance hosted in Cloud@CNAF is used for storage instead of the local file system. Details on the aspects related to distributed computing are discussed in Section~\ref{sec:interlink}.

If run sequentially on a single node, the workflow takes O(10 hours) per simulated geometry to complete. Parallelization is achieved by simulating different traversing pions in parallel instances and merging the results before introducing the effects of the electronics and of the DAQ system.

\section{GPU-Accelerated Simulation with \emph{Garfield++} and \emph{WeightingTide}}
\label{sec:weightingtide}
The numerical methods to compute the electrostatic field solving the Laplace equation and determine the correlation function $\vec H(\vec x, t)$ by solving the Maxwell equations in the quasi-static approximation are described in Ref.~\cite{Anderlini:2025ffy}. In this Section, we summarize the overall strategy of the methods and focus on the aspects relevant to their implementation on GPU. 

The simulation was implemented in Python using Cupy, a Python package exposing API very similar to those of Numpy to GPU-accelerated kernels, possibly interleaved with custom CUDA kernel defined in the Python script and compiled ahead of time.

\begin{figure}
    \centering
    \begin{minipage}[t]{0.4\textwidth}
        \includegraphics[width=\textwidth]{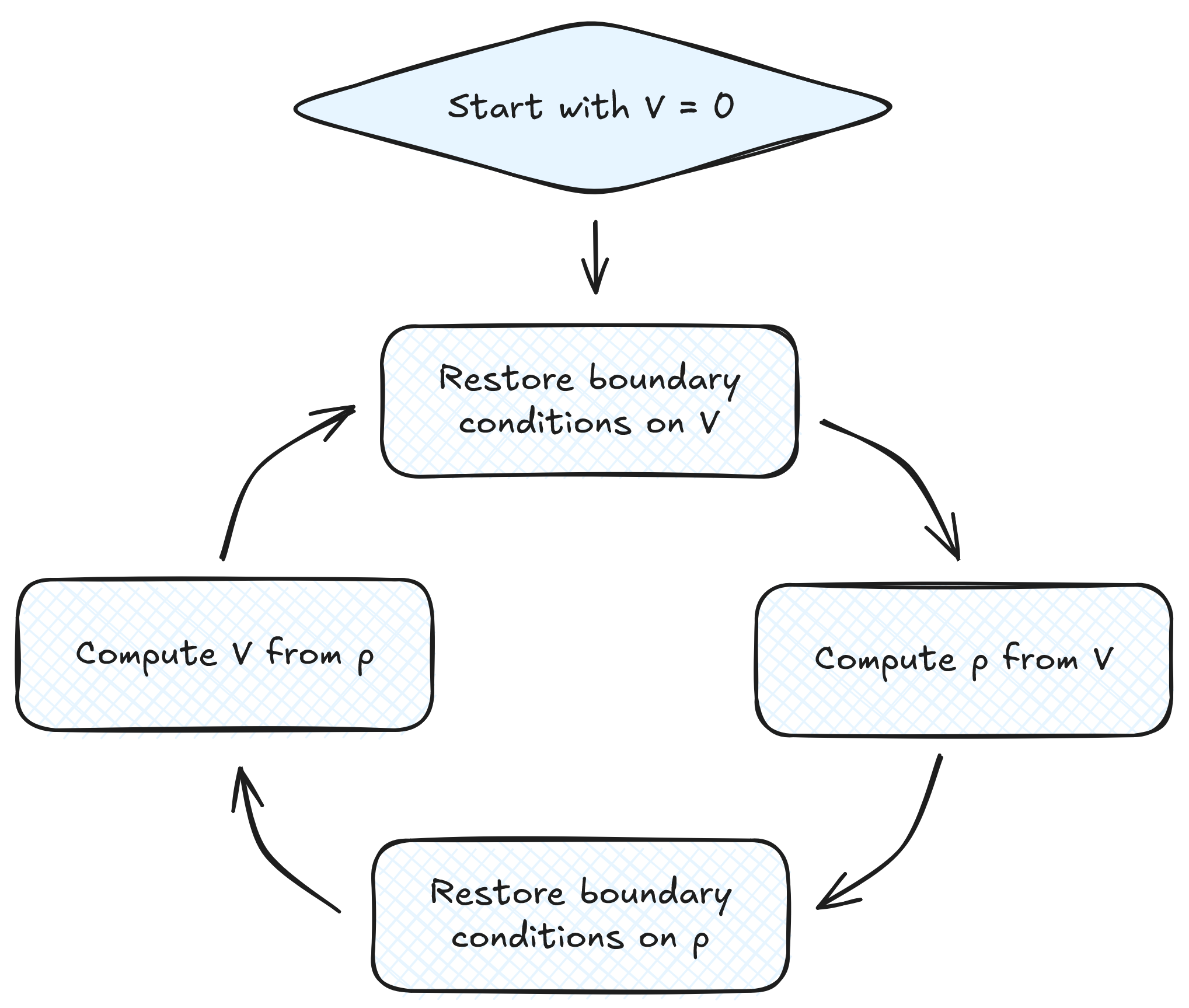}
        \caption{\label{fig:wt-static-maps} Iterative procedure used to solve the Laplace equation in \emph{WeightingTide} implementation.}
    \end{minipage}
    \hfill
    \begin{minipage}[t]{0.58\textwidth}
        \includegraphics[width=\textwidth]{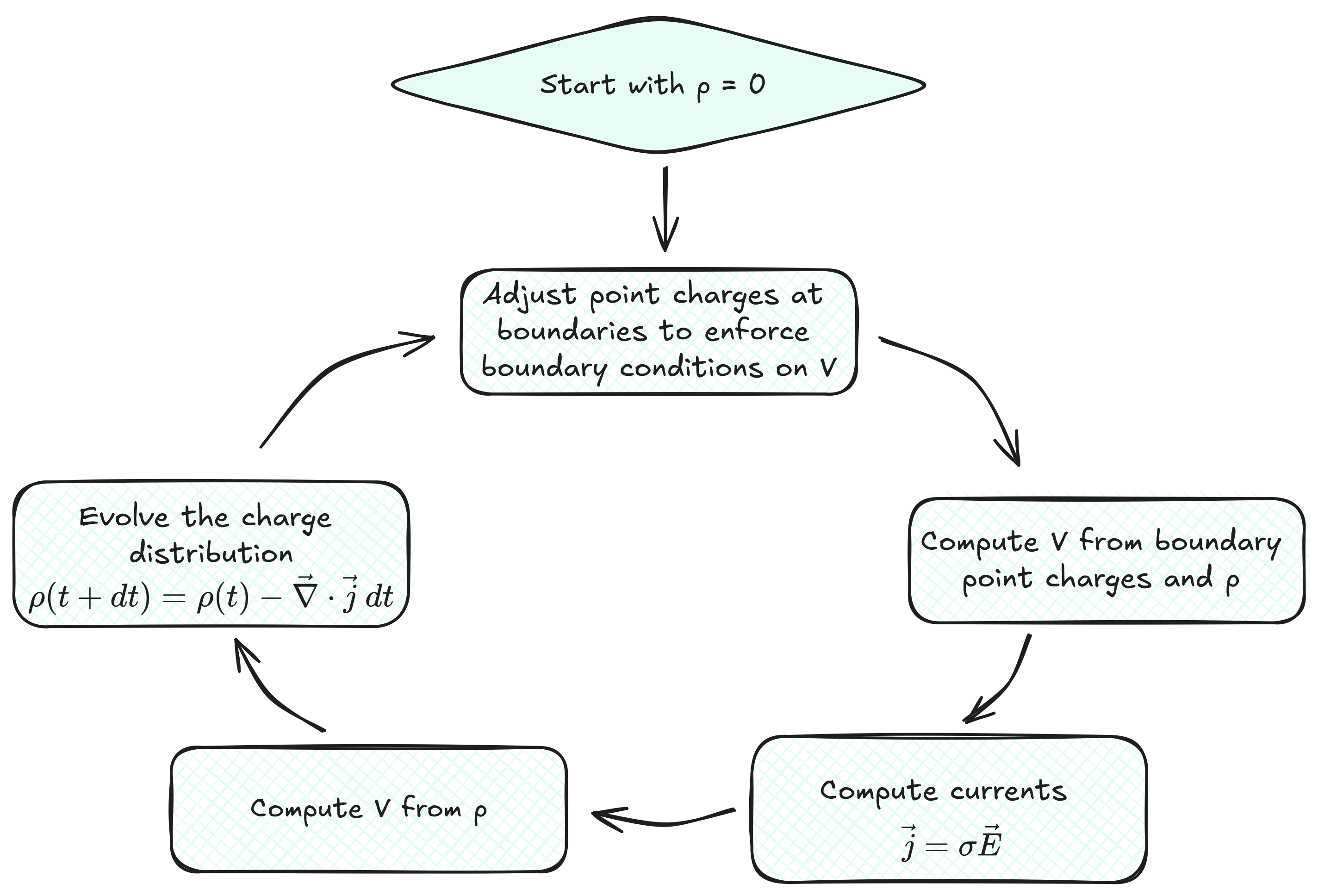}
        \caption{\label{fig:wt-dynamic-maps} Iterative procedure used to solve the Maxwell equations in the quasi-static approximation in \emph{WeightingTide} implementation.}
    \end{minipage}
\end{figure}

In particular, in the iterative procedure used to compute the electrostatic field, depicted in Figure~\ref{fig:wt-static-maps}, the computation of the charge distribution $\rho$ from the potential $V$ (and vice-versa) is obtained combining the 3D FFT implementation shipped with CUDA with a custom kernel multiplying (or dividing) the resulting spectral representation with the squared momentum of the wave-vector $|\vec k|^2$. Using the custom kernel instead of a Python implementation halves the GPU memory required to perform the multiplication.

Similarly, in the iterative forward-Euler procedure used to compute the correlation function $\vec H(\vec x, t)$, a custom CUDA kernel was developed to compute the voltage set by the point charges placed at the boundaries of the active volume and adjusted, in magnitude, to enforce the boundary conditions. This operation involves intensive memory operations to copy rescaled and translated chunks of the fundamental solution to array representing the external contribution to the potential (indicated as $V_{\mathrm{ext}}$ in Ref.~\cite{Anderlini:2025ffy}).
Since memory access to perform this step dominated the CPU implementation, the $100\times$ speedup achieved porting the computation of the correlation function to GPU is not surprising.

The most extensive use of custom CUDA kernels was done for the convolution of the carrier velocity as inferred by the trajectories computed by Garfield++ and the correlation function $\vec H(\vec x, t)$. In this case, we chose to define the whole correlation for a single carrier in a custom CUDA kernel and let the GPU schedule the convolution of the $O(10\,000)$ carriers in parallel threads. Further speed-up was achieved by loading the correlation function $H$ into the texture memory of the GPU to profit from hardware implementation of the linear interpolation along the carrier's trajectory. 

While the exact speed-up depends on hardware and geometry, GPU acceleration reduced time-to-insight from one week to a few hours. At the time of writing, computing the paths of the carriers in Garfield++ dominates the time-to-insight, motivating further optimization of the step size used to integrate the electrical field.

\section{Distributed simulation throughout the TeRABIT project with InterLink}
\label{sec:interlink}

Porting the computation heavy software on GPU in a consistent and stable workflow enables, at least conceptually, scans and optimizations by submitting several configurations at once and then studying how the resolution depends on geometrical parameters. Practically, however, running tens of instances of a multiple-hour workflow sequentially, would still result in unacceptably long time-to-insight. Hence distribution.

In the context of the ICSC and TeRABIT projects, our research was granted access to the HTC resources of the CNAF WLCG Tier-1 site, CPU-only nodes managed with HTCondor, and to the HPC site in Padova, providing several H200 GPUs managed in a SLURM~\footnote{see \href{https://slurm.schedmd.com}{slurm.schedmd.com}} cluster.  

The Snakemake workflow, built and debugged collaboratively on the Cloud resources managed by the AI\_INFN initiative (INFN CSN5), was then configured to launch the jobs towards the underlying Kubernetes cluster. Kubernetes jobs are queued using Kueue~\footnote{see \href{https://kueue.sigs.k8s.io/}{kueue.sigs.k8s.io}}. Kueue is configured to assign the jobs to different \emph{resource flavors} based on the requested resources and on the availability of the remote computing sites, and then to schedule the jobs on Kubernetes nodes corresponding to the assigned \emph{resource flavor}. In the AI\_INFN Kubernetes cluster, the InterLink project is used to define virtual nodes that execute the containerized computing payloads in remote sites rather than in a physical underlying computing node. Kueue was configured to match the queue for this research to either local nodes or to the virtual nodes corresponding to the CNAF Tier-1 site and to the Padova HPC bubble. 

The payloads distributed to geographically distant sites exchange data accessing to a centralized S3 storage, provided by a Deuxfleur Garage instance in the Cloud region of CNAF.
The access to the storage is the main bottleneck in the parallel execution of the workflow as the data transfer bandwidth gets saturated before the availability of computing resources is exceeded.

Software is distributed through the CERN Virtual Machine File System (CVMFS) as containerized computing environment, managed by Apptainer. CVMFS Unpacked was used to transparently propagate the images to the INFN sites involved and cache them on the computing nodes. 

\begin{figure}
    \centering
    \includegraphics[width=\textwidth]{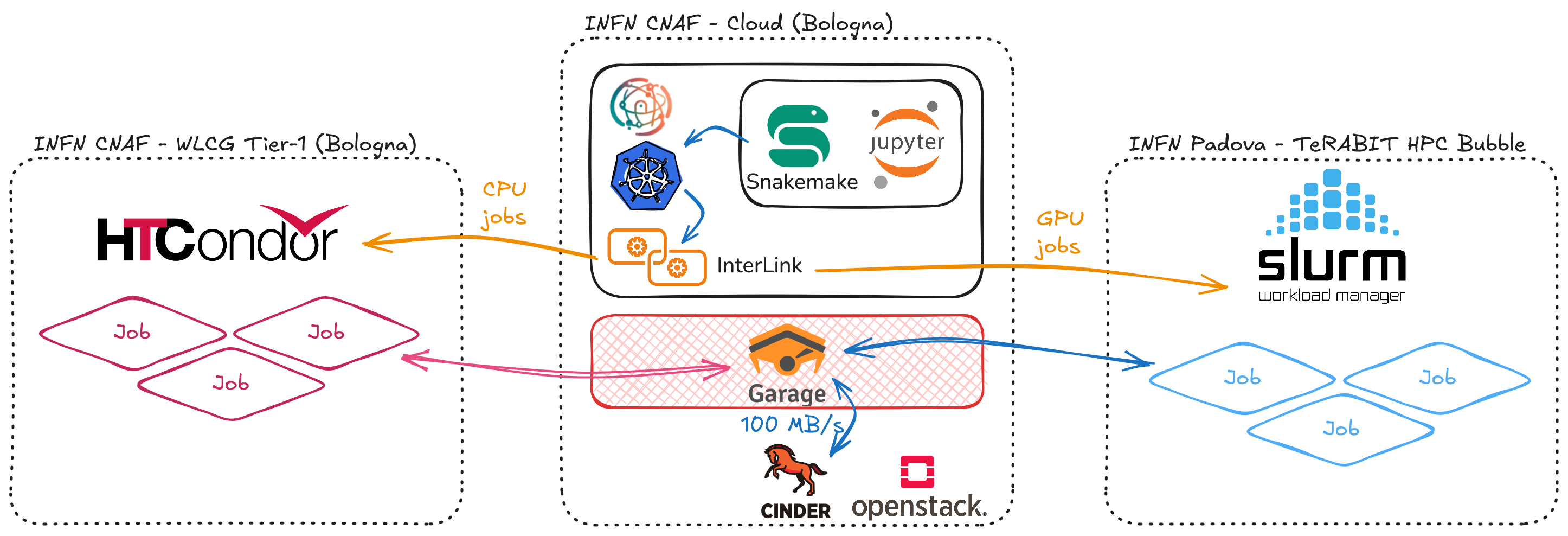}
    \centering
    \caption{\label{fig:ai-infn-flow} 
        Distribution of the computing payloads from a Kubernetes cluster in Bologna to 
        the CNAF WLCG Tier-1 site managed with HTCondor and one of the HPC Bubbles 
        of the TeRABIT Project in Padova.
    }
\end{figure}

\section{Results}
\label{sec:results}
\begin{figure}
    \centering
    \begin{minipage}{0.4\textwidth}
        \includegraphics[width=\textwidth]{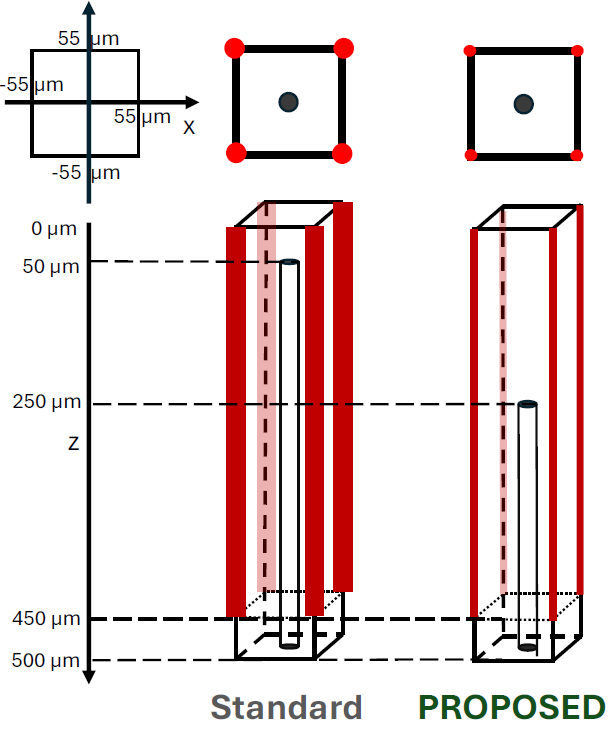}
    \end{minipage}
    \hfill
    \begin{minipage}{0.4\textwidth}
       \caption{\label{fig:proposedgeometry}
        On the left, a pixel representing the standard geometry used for the 2021 test beam detector; on the right, the geometry proposed following the simulation studies.
        } 
    \end{minipage}
\end{figure}

Currently, the state-of-the-art time resolution for this type of detector is well below 100 ps, achieved during a beam test at the CERN SPS in 2021~\cite{ANDERLINI2022167230}. The tested device consisted of a prototype $6\times6$ 3D pixel matrix with a pitch of 55$\times$55~$\mu\mathrm{m}^2$, a schematic view of a standard single pixel can be seen in Figure~\ref{fig:proposedgeometry}. The electrode diameter was approximately 12~$\mu\mathrm{m}$, with a resistance of about 30~$\mathrm{k}\Omega$ (equivalent to a resistivity of 0.75~$\Omega\cdot$cm), and a bulk thickness of 500~$\mu\mathrm{m}$. The simulation was validated using the data from this test beam~\cite{Anderlini:2025ffy}.

Thanks to the new simulation architecture, it is now possible to carry out extensive parametric studies of the performance of 3D diamond detectors with graphitized electrode. We focused our investigation on two main geometrical aspects aimed at improving the time resolution while facilitating sensor fabrication. Starting from the baseline geometry of the test beam detector, we independently studied the effect of the bias electrode diameter and the diamond bulk thickness.

\begin{figure}
    \centering
    \begin{minipage}[t]{0.49\textwidth}
        \includegraphics[width=\textwidth]{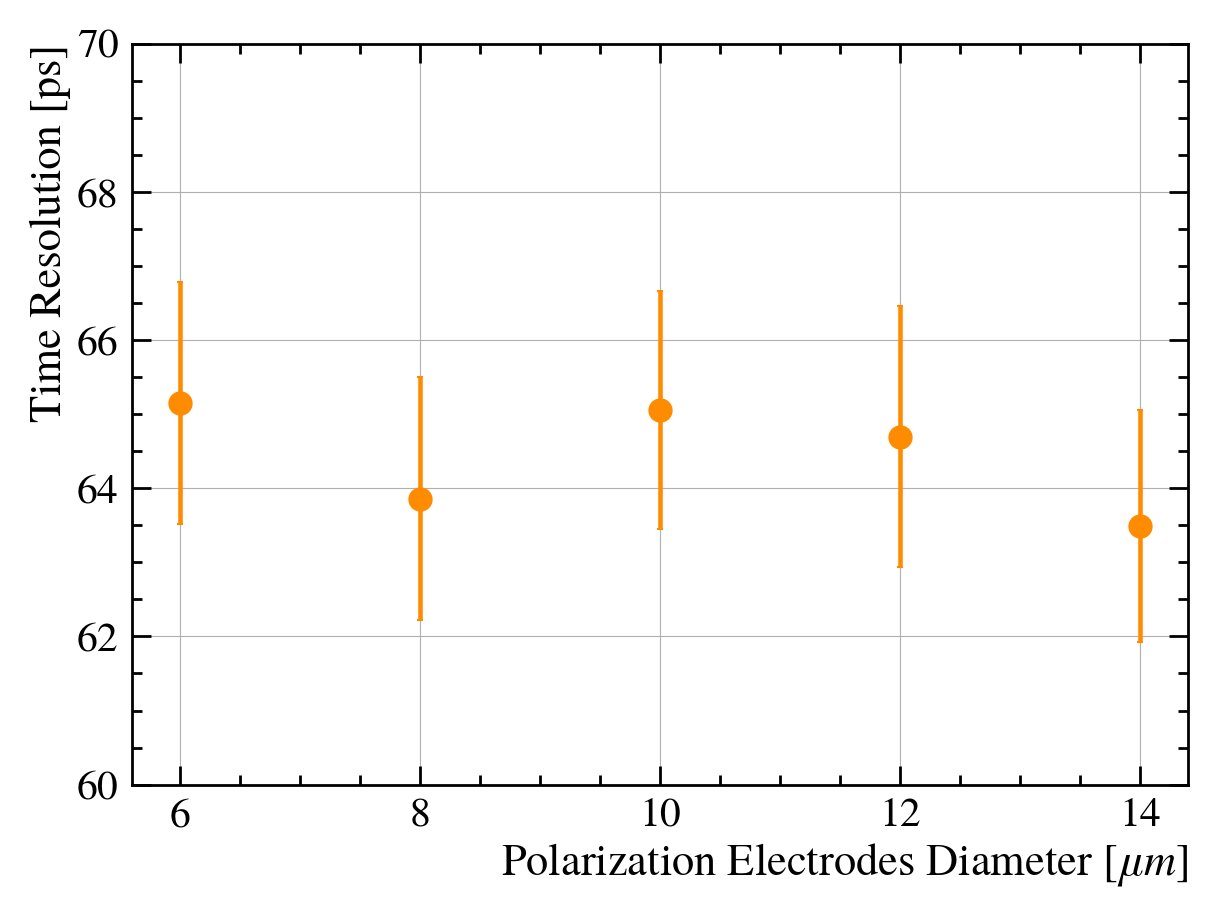}
        \caption{\label{fig:diam} 
        Time resolution as a function of bias electrodes diameter.
        }
    \end{minipage}
    \hfill
    \begin{minipage}[t]{0.49\textwidth}
        \includegraphics[width=\textwidth]{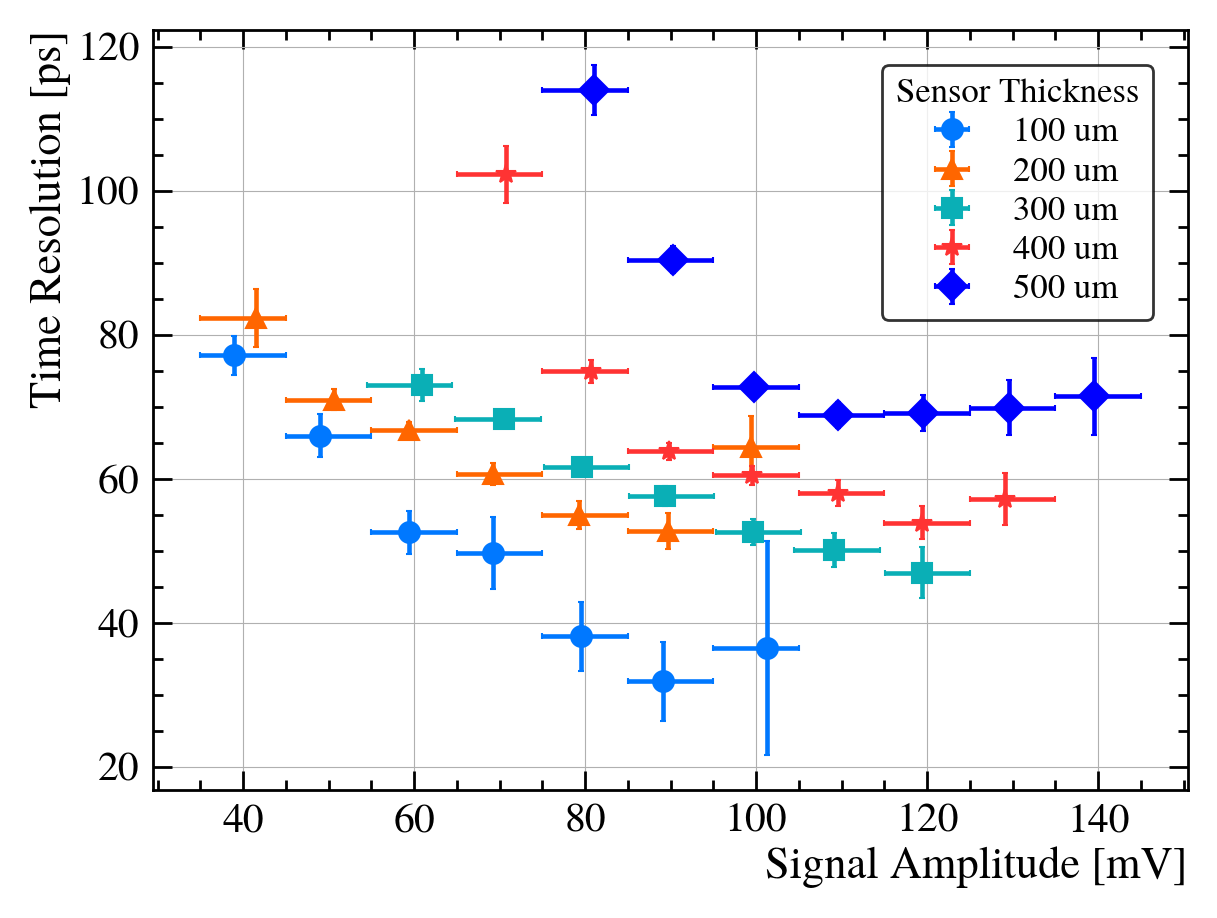}
        \caption{\label{fig:bulk} 
        Time resolution as a function of the signal amplitude for different sensor thickness.
        }
    \end{minipage}
\end{figure}

Figure~\ref{fig:diam} shows the time resolution as a function of the bias electrode diameter $d$. Modifying the electrode diameter directly affects its resistance $R = 4\rho L/(\pi d^2)$, where $\rho$ is the material resistivity and $L$ is the electrode length. The bias electrode diameter was varied from 6~$\mu\mathrm{m}$ to 14~$\mu\mathrm{m}$ in steps of 2~$\mu\mathrm{m}$.
The detector time resolution does not exhibit a significant dependence on the bias electrode diameter, and consequently, on its resistance. At first glance, this might seem contradictory, as it has been established that the primary contribution to the time resolution in these detectors is the electrode resistance. However, it is crucial to distinguish which electrodes are involved.
In Equation~\ref{eq:ramoshockleyriegler}, the resistance contribution is embedded in the time dependence of the correlation function $\vec H(\vec x, t)$, which depends on all impedances within the sensor and is derived from a differential equation with specific boundary conditions, discussed in detail in~\cite{Anderlini:2025ffy}. Any variation in the boundary conditions affects the induced current in Equation~\ref{eq:ramoshockleyriegler} through the evolution of $\vec H(\vec x, t)$. In our case, the 3D geometry makes the contribution from the bias electrodes practically negligible. Their initial and final conditions are nearly identical, resulting in a minimal impact on signal formation and therefore on time resolution. Instead, the resistance of the readout electrodes plays a dominant role, as it directly affects signal propagation and shaping.
This result indicates an opportunity to reduce the risk of damaging the diamond crystal during the graphitization process by fabricating bias electrodes with very small diameters.

Figure~\ref{fig:bulk} illustrates the time resolution as a function of signal amplitude for various diamond thicknesses, effectively probing the impact of different readout electrode lengths. Time resolution follows a typical trend: as amplitude increases, the noise contribution decreases, leading to a minimum asymptotic resolution characteristic of the specific sensor. 
Interestingly, for thinner diamond bulks, the asymptotic time resolution decreases. This is consistent with the fact that the dominant contribution to the time resolution in these detectors arises from signal propagation along the readout electrode. Therefore, reducing the length of the readout electrode directly enhances timing performance.
Based on these two studies, we propose two initial geometric optimizations that have minimal impact on the fabrication process: shortening the readout electrodes, balancing this with signal amplitude to maintain efficiency, and reducing the diameter of the bias electrodes to the fabrication limit, as illustrated in Fig.~\ref{fig:proposedgeometry}. These modifications are expected to improve the time resolution by approximately 10$\%$, assuming uniform conductivity within the electrodes. This assumption may be conservative, as a depth-dependent conductivity profile could result in an even greater improvement.

\section{Conclusion and Outlook}
\label{sec:conclusion}

Consistent with previous studies, improving the conductivity of the graphitized columns remains the most effective strategy for further enhancing the performance of diamond sensors. As incremental gains of about 10$\%$ are progressively achieved through sensor design optimization, the detector performance is rapidly approaching the fundamental limits imposed by the readout electronics. Overall, these results identify a promising path toward achieving state-of-the-art time resolution in 3D diamond detectors.

\acknowledgments

This work was made possible by the support of INFN DataCloud and of the AI\_INFN initiative. We wish to thank in particular C. Pellegrino and A. Pascolini for supporting access to the CNAF Tier-1 resources, M. Sgaravatto and M. Verlato for supporting access to the TeRABIT resources in INFN Padova, and D. Ciangottini and G. Bianchini for supporting the configuration of InterLink.  


\bibliographystyle{JHEP}
\bibliography{biblio.bib}

\end{document}